\documentclass[11pt]{article}
\usepackage{jheppub}

\usepackage{amsmath}

\newcommand{\ben}[1]{\begin{eqnarray}\label{#1} }
\newcommand{\een}{\end{eqnarray}}

\newcommand{\refb}[1]{(\ref{#1})}

\newcommand{\DD}{{\cal D}}

\newcommand{\thc}{\text{h.c.}}
\newcommand{\Dslash}{\not{\hbox{\kern-4pt $D$}}}
\newcommand{\pslash}{\not{\hbox{\kern-4pt $\partial$}}}
\newcommand{\Dcslash}{\not{\hbox{\kern-4pt $\DD$}}}

\usepackage{graphicx}
\usepackage{amssymb}
\usepackage{amsthm}
\usepackage{amsmath}
\usepackage{slashed}
\usepackage{color}
\usepackage{bbold}

\newcommand{\gU}{{\rm U}}
\newcommand{\gSU}{{\rm SU}}

\numberwithin{equation}{section}

\newcommand {\cB}{{\cal B}}
\newcommand {\cC}{{\cal C}}

\newcommand {\cF}{{\cal F}}
\newcommand {\cG}{{\cal G}}
\newcommand {\cH}{{\cal H}}

\newcommand {\cJ}{{\cal J}}

\newcommand {\cW}{{\cal W}}
\newcommand {\cX}{{\cal X}}

\def\a{\alpha}
\def\b{\beta}

\def\g{\gamma}
\def\G{\Gamma}

\def\q{\theta}

\def\L{\Lambda}

\newcommand{\rd}{{\mathrm d}}

\newcommand{\ad}{{\dot{\alpha}}}
\newcommand{\bd}{{\dot{\beta}}}
\newcommand{\dalpha}{{\dot{\alpha}}}
\newcommand{\dbeta}{{\dot{\beta}}}
\newcommand{\dgamma}{{\dot{\gamma}}}

\newcommand{\veps}{\varepsilon}
\newcommand{\eps}{{\epsilon}}
\newcommand{\eol}{\notag \\}

\newcommand{\ul}{\underline}

\makeatletter
\g@addto@macro\bfseries{\boldmath}
\makeatother

\title{$N=2$ dilaton Weyl multiplet in 4D supergravity}

\author[a]{Daniel Butter,}
\author[b]{Subramanya Hegde,}
\author[c,d]{Ivano Lodato,}
\author[b]{and Bindusar Sahoo}

\affiliation[a]{
George P. and Cynthia Woods
Mitchell Institute for 
Fundamental Physics and Astronomy, 
Texas A\&{}M University,
College Station, TX 77843, USA}

\affiliation[b]{
Indian Institute of Science Education and Research,
Vithura, Thiruvananthapuram - 695551, India}

\affiliation[c]{
Indian Institute of Science Education and Research,
Homi Bhabha Road, Pashan, Pune 411 008, India }

\affiliation[d]{
Department of Physics and Center for Field Theory and Particle Physics,
Fudan University,
220 Handan Road, 200433 Shanghai, China
}

\emailAdd{dbutter@tamu.edu}
\emailAdd{smhegde14@iisertvm.ac.in}
\emailAdd{ilodato@fudan.edu.cn}
\emailAdd{bsahoo@iisertvm.ac.in}

\arxivnumber{1712.xxxxx}

\abstract{We construct the dilaton Weyl multiplet for $N=2$ conformal supergravity in four dimensions. Beginning from an on-shell vector multiplet coupled to the standard Weyl multiplet, the equations of motion can be used to eliminate the supergravity auxiliary fields, following a similar pattern as in five and six dimensions. The resulting 24+24 component multiplet includes two gauge vectors and a gauge two-form and provides a variant formulation of $N=2$ conformal supergravity.
We also show how this dilaton Weyl multiplet is contained in the minimal 32+32 Poincar\'e supergravity multiplet introduced by M\"uller \cite{Muller:1986ku} in superspace.
}

\begin{document}
\allowdisplaybreaks
\maketitle

\section{Introduction}

The superconformal multiplet calculus is a crucial tool for the construction and analysis of general matter couplings to gauge theories, both in flat and curved space. Though conformality implies the absence of a physical energy scale, necessary for both a classical as well as quantum description, superconformal invariant couplings of supergravity can be gauge fixed to obtain the physical Poincar\'e theory. 
This procedure requires compensator multiplets, short matter multiplets which are used to fix conformal symmetries.
The knowledge of general irreducible representations of the superconformal algebra is hence of utmost importance to the construction of invariant couplings in supergravity, especially higher derivative ones,
see e.g.
\cite{Bergshoeff:1980is,
deWit:2010za,
Butter:2013lta,
Kuzenko:2015jxa,
Hanaki:2006pj,
Ozkan:2013nwa,
Butter:2014xxa},
which are relevant to our discussion here.

Off-shell superconformal multiplets with 8 supercharges exist for $3 \!\leq d \!\leq 6$.
This includes the Weyl multiplet of conformal supergravity itself, which will be
our chief concern in this paper.
These Weyl multiplets in differing spacetime dimensions
are deeply connected and, in fact, the various cases can be connected
via dimensional reduction \cite{Fujita:2001kv,Banerjee:2011ts,Banerjee:2015uee}.
This entails that certain field representations and structures in $d$ dimensions
must correspond to those in $d-1$ dimensions.
One important example of such correspondence, central to this paper, is the existence of
a variant version of the Weyl multiplet, with the same number of off-shell degrees of
freedom, known as the dilaton Weyl multiplet.

The dilaton Weyl multiplet was discovered first in six dimensions in \cite{Bergshoeff:1985mz},
and later on in five dimensions \cite{Bergshoeff:2001hc}. 
As the name suggests, these variant Weyl multiplets contain a scalar field of
non-vanishing Weyl weight. Because the off-shell degrees of freedom match the corresponding
standard Weyl multiplet, a bosonic degree of freedom from the standard Weyl multiplet must be
removed in exchange for the dilaton. This field is an auxiliary scalar of Weyl weight two,
typically denoted $D$.
This is not the only exchange which occurs. Supersymmetry implies a dilatino partner
to the dilaton, and this is exchanged with a fermionic auxiliary $\chi^i$. Finally,
an auxiliary tensor field is exchanged for a set of $p$-forms. In six dimensions,
an anti-self-dual tensor $T_{abc}^-$ is traded for a dynamical gauge two-form $B_{\mu\nu}$,
while in five dimensions, an antisymmetric tensor $T_{ab}$ is traded for a gauge
two-form $B_{\mu\nu}$ as well as a gauge one-form $A_\mu$.

This exchange is helpful for a few reasons. First, in two-derivative actions,
the auxiliary $D$ plays the role of a Lagrange multiplier and so there must
be another scalar degree of freedom for it to remove to construct a physical action
(and similarly with $\chi^i$).
This is the reason that theories with eight supercharges require two
compensators: one scalar degree of freedom becomes a Weyl compensator whose kinetic
term yields the Einstein action, while another must be paired with $D$.
In the dilaton Weyl multiplet, only one compensating multiplet is needed to provide
the Einstein action.
Another advantage of the 5d and 6d dilaton Weyl multiplets is that they both come with
an off-shell two-form built in. In the standard Weyl multiplets, any two-form must come
from a superconformal matter multiplet, but in both $d=5$ and $d=6$, the short matter
multiplets involving two-forms are on-shell.\footnote{Actually, in five dimensions,
the off-shell two-form multiplet can be defined in the presence of a central charge \cite{Kugo:2002vc} (see \cite{Kuzenko:2006mv,Kuzenko:2013rna} for its superspace form),
consistent with its reduction from six dimensions.} This suggests that constructing general
higher derivative actions, such as those descending from string theory,
would be difficult.
The dilaton Weyl multiplet naturally provides a means to introduce a single
two-form into the gravity multiplet (with the dilaton and dilatino as an added bonus)
while remaining off-shell.

It has long been suspected that a $d=4$ dilaton Weyl multiplet should be possible,
but, to the best of our knowledge, it has not been explicitly constructed.
It was first noticed by Siegel \cite{Siegel:1995px} that a variant Weyl multiplet
in $d=4$ was possible via a string-inspired argument. By formally tensoring
together two copies of (super) Yang-Mills from the left and right-moving modes
of a string compactified to four dimensions, Siegel derived the field content
of the standard Weyl multiplet and the dilaton Weyl multiplet from, respectively,
an $N=(1,1)$ worldsheet and an $N=(2,0)$ worldsheet (with the left-moving sector
of type II). The $d=4$ dilaton Weyl multiplet differs from
its higher dimensional analogues with the auxiliary anti-self-dual tensor $T_{ab}^-$
replaced by a gauge two-form and \emph{two} gauge one-forms. As with $d=5$, this
set of $p$-forms is precisely what one would expect from dimensionally reducing
the $d=6$ two-form.
In fact, this $d=4$ dilaton Weyl multiplet was \emph{almost} constructed already in superspace
even earlier by M\"uller \cite{Muller:1986ku}. He considered a larger
$32+32$ Poincar\'e supergravity multiplet, which he called the new minimal $N=2$ supergravity
multiplet, which it turns out can be interpreted as the dilaton Weyl
multiplet coupled to a linear multiplet compensator.

In this paper, we present a modern analysis, inspired by the five dimensional case \cite{Bergshoeff:2001hc}. There the dilaton Weyl multiplet was found in two different, yet equivalent ways: via the construction of the current multiplet for a non-conformal vector multiplet and by coupling the standard Weyl multiplet to an on-shell vector multiplet.
The two methods are in fact not equivalent in four dimensions. Coupling to an on-shell
vector multiplet leads to the actual dilaton Weyl multiplet, while the current analysis
leads instead to a large irreducible 24+24 component matter multiplet.
Here we focus on the dilaton Weyl multiplet, leaving the latter discussion to
another work \cite{Hegde:2017sgl}.
In section 2, we review the standard Weyl multiplet, with some minor modifications
to its conventional constraints to separate out the auxiliary fields $D$, $\chi^i$,
and $T_{ab}^-$. In section 3, we analyze the on-shell vector multiplet and solve its
multiplet of field equations. The dilaton Weyl multiplet is summarized in section 4.
In section 5, we show how to modify M\"uller's construction \cite{Muller:1986ku}
to remove the linear multiplet compensator, giving a superspace description of the
dilaton Weyl multiplet.

\section{The standard Weyl multiplet and its (un)conventional constraints}
The structure of the $N=2$ standard Weyl multiplet was worked out in \cite{deWit:1979dzm} several decades ago. It is an irreducible field representation of the superconformal algebra $\gSU(2,2|2)$ consisting of the following field content. There is a vierbein $e_\mu{}^a$, a gravitino $\psi_\mu{}^i$, connections $V_\mu{}^i{}_j$ and $A_\mu$ gauging the $R$-symmetry group $\gSU(2)_R \times \gU(1)_R$, and a dilatation connection $b_\mu$. The remaining gauge fields are the Lorentz spin connection $\omega_\mu{}^{ab}$, the $S$-supersymmetry connection $\phi_\mu{}^i$, and the special conformal (or K-boost) connection $f_\mu{}^a$. These latter three connections turn out to be composite and determined in terms of the so-called \emph{conventional constraints}
\begin{align}\label{E:ConventionalConstraints}
R(P)_{\mu\nu}{}^a=0\;,\qquad \g^\mu\,R(Q)_{\mu\nu}{}^i+\tfrac32\,\g_\nu\,\chi^i=0\;,
\nonumber\\
e^\mu{}_b\,R(M)_{\mu\nu a}{}^b+{\rm i}\,\tilde{R}(A)_{\mu a}+\tfrac18 T_{abij}\,T_\mu{}^{bij}-\tfrac32\,D e_{\mu a}=0~.
\end{align}  
Here $R(P)$, $R(M)$, $R(A)$ and $R(Q)$ are the supercovariant curvatures associated with local translations, local Lorentz transformations, chiral $\gU(1)_R$ transformations, and the ordinary supersymmetry (also referred to as Q-supersymmetry). The covariant fields $T_{ab}{}^{ij}$, $D$ and $\chi^{i}$ appearing both explicitly in these constraints and implicitly in the definitions of the curvatures themselves are necessary for completing the field representation. In all two-derivative actions, these covariant fields appear without kinetic terms and are generally called the \emph{auxiliary fields} of the Weyl multiplet for this reason. These auxiliary fields ensure that the off-shell bosonic and fermionic degrees of freedom match, providing $7+8$ degrees of freedom to yield the total $24+24$ off-shell degrees of freedom of the standard Weyl multiplet.

In five and six dimensions, the standard Weyl multiplet with eight supercharges shares a very similar structure, including a similar auxiliary field content. There it is possible to introduce  an alternative Weyl multiplet, known as the dilaton Weyl multiplet, where these auxiliary fields are removed. In their place, one introduces a dynamical scalar field (the dilaton), a dynamical fermion (the dilatino), and a set of $p$-form gauge fields that lead to a new multiplet of $24+24$ off-shell degrees of freedom. In both cases, the dilaton Weyl multiplet can be viewed as the standard Weyl multiplet coupled to a certain on-shell multiplet (a vector multiplet for $d=5$ and a tensor multiplet for $d=6$). The field equations for the on-shell multiplet are linear in the Weyl multiplet auxiliaries and can be reinterpreted as constraint equations for the auxiliaries themselves. This exchanges the Weyl multiplet auxiliaries for these other physical fields, while keeping the total number of off-shell degrees of freedom unchanged.

To construct the 4d dilaton Weyl multiplet, we will follow the same procedure used in in five dimensions \cite{Bergshoeff:2001hc}: we couple an $N=2$ vector multiplet to the standard Weyl multiplet and use the equations of motion to eliminate the auxiliary fields. To do this efficiently, it is helpful to extract from the composite gauge fields $\phi_{\mu i}$ and $f_{\mu}{}^{a}$ any dependence on the auxiliary fields $T_{ab}{}^{ij}$, $\chi^{i}$, and $D$. That is, one wishes to shift the auxiliary fields as
\begin{align}\label{changes}
\phi_{\mu i}^{\text{(new)}}&=\phi_{\mu i}^{\text{(old)}}-\frac{1}{2}\gamma_{\mu}\chi_{i}~, \nonumber\\
f_{\mu a}^{\text{(new)}}&= f_{\mu a}^{\text{(old)}}+\frac{1}{4}De_{\mu a}-\frac{1}{16}T_{\mu b}{}^{ij}T_{a}{}^{b}{}_{ij}\;.
\end{align}
These redefinitions actually correspond to making a different choice of conventional
constraints
\begin{align}\label{unconven_constraints}
R(P)_{\mu\nu}^{a}=0~, \qquad
\gamma^{\mu}R(Q)_{\mu\nu i} =0~, \qquad
e_{b}{}^{\nu}R(M)_{\mu\nu}{}_{a}{}^{b}-i\tilde{R}(A)_{\mu a}=0~.
\end{align}
Comparing with \eqref{E:ConventionalConstraints}, these (un)conventional constraints
correspond to curvature shifts
\begin{align}
R(Q)_{\mu\nu i}^{\text{(new)}}&=R(Q)_{\mu\nu i}^{\text{(old)}}+\frac{1}{2}\gamma_{\mu\nu}\chi_{i}~,\nonumber \\
R(M)^{\text{(new)}}_{\mu\nu}{}^{ab}&=R(M)^{\text{(old)}}_{\mu\nu}{}^{ab}-De_{[\mu}{}^{a}e_{\nu]}{}^{b}+\frac{1}{4}T_{[\mu|c}{}^{ij}T^{[a|c}{}_{ij}e_{|\nu]}{}^{|b]}~.
\end{align}
The transformations under Q- and S-supersymmetry will also change accordingly,
\begin{align}\label{WeylTransf}
\delta e_{\mu}{}^{a} &= \bar{\epsilon}^{i}\gamma^{a}\psi_{\mu i}+\thc\nonumber \\
\delta\psi_{\mu}{}^{i} &= 2\,\mathcal{D}_{\mu}\epsilon^{i}-\frac{1}{8}\gamma \cdot T^{ij}\gamma_{\mu}\epsilon_{j}-\gamma_{\mu}\eta^{i}\nonumber \\
\delta b_{\mu}&=\frac{1}{2}\bar{\epsilon}^{i}\phi_{\mu i}-\frac{1}{2}\bar{\epsilon}^{i}\gamma_{\mu}\chi_{i}-\frac{1}{2}\bar{\eta}^{i}\psi_{\mu i}+\thc+\Lambda_{K}^{a}e_{\mu a} \nonumber \\
\delta A_{\mu} &= \frac{i}{2}\bar{\epsilon}^{i}\phi_{\mu i}+i\bar{\epsilon}^{i}\gamma_{\mu}\chi_{i}+\frac{i}{2}\bar{\eta}^{i}\psi_{\mu i}+\thc \nonumber \\
\delta V_{\mu}{}^{i}{}_{j} &= 2\,\bar{\epsilon}_{j}\phi_{\mu}^{i} - 2\,\bar{\epsilon}_{j}\gamma_{\mu}\chi^{i} + 2\,\bar{\eta}_{j}\psi_{\mu}^{i} - (\thc;~\text{traceless})\nonumber \\
\delta T_{ab}{}^{ij}&=8\,\bar{\epsilon}^{[i}R(Q)_{ab}{}^{j]}-4\bar{\epsilon}^{[i}\gamma_{ab}\chi^{j]}\nonumber \\
\delta\chi^{i}&=-\frac{1}{12}\gamma_{ab}\Dslash T^{abij}\epsilon_{j} + \frac{1}{6} \gamma \cdot R(V)^{i}{}_{j}\epsilon^{j} - \frac{i}{3}\gamma \cdot R(A)\epsilon^{i}+D\epsilon^{i}+\frac{1}{12}\gamma\cdot T^{ij}\eta_{j}\nonumber \\
\delta\omega_{\mu}{}^{ab} &= 
	-\frac{1}{2}\bar{\epsilon}^{i}\gamma^{ab}\phi_{\mu i}
	-\frac{1}{4}\bar{\epsilon}^{i}[\gamma^{ab},\gamma_{\mu}]\chi_{i}
	-\frac{1}{2}\bar{\epsilon}^{i}T^{ab}{}_{ij}\psi_{\mu}^{j}
	+\bar{\epsilon}^{i}\gamma_{\mu}R(Q)^{ab}{}_{i}
	\nonumber \\ & \quad
	-\frac{1}{2}\bar{\eta}^{i}\gamma^{ab}\psi_{\mu i}
	+\thc
	+2\Lambda_{K}^{[a}e_{\mu}{}^{b]}\nonumber\\
\delta^{\text{cov}}\phi_{\mu}{}^{i} &=
	-\frac{1}{8}\Big[\Dslash \gamma\cdot T^{ij}\gamma_{\mu}
	-\frac{1}{3}\gamma_{\mu}\gamma\cdot\Dslash T^{ij}\Big]\epsilon_{j}
	+\frac{1}{4}\Big[\gamma\cdot R(V)^{i}{}_{j}\gamma_{\mu}
	-\frac{1}{3}\gamma_{\mu}\gamma\cdot R(V)^{i}{}_{j}\Big]\epsilon^{j}
	\nonumber \\ & \quad
	+ \frac{i}{2}\Big[\gamma\cdot R(A)\gamma_{\mu}+\frac{1}{3}\gamma_{\mu}\gamma\cdot R(A)\Big]\epsilon^{i}
	-\frac{1}{8}T_{\mu b}{}^{kj}T^{ab}{}_{kj}\gamma_{a}\epsilon^{i}
	-\frac{1}{24}\gamma_{\mu}\gamma\cdot T^{ij}\eta_{j}~.
\end{align}
Here we have refrained from giving the transformation rule for $f_{\mu}{}^{a}$ and the non-covariant pieces in the transformation law of $\phi_{\mu}^{i}$. They can be easily worked out, but are not relevant for our computation. It is a simple exercise to verify that all the fields appearing above obey the new constraints.

In the presence of the modified constraints (\ref{unconven_constraints}), the supersymmetry algebra is altered. One finds that, as with the conventional constraints, the superconformal algebra of $\gSU(2,2|2)$ must be modified, introducing new field-dependent structure constants
(or \emph{structure functions}). The supersymmetry algebra becomes
\begin{align}\label{algebra}
\left[\delta_Q(\epsilon_1),\delta_Q(\epsilon_2)\right]&=\delta^{(cov)}(\xi)+\delta_{M}(\varepsilon)+\delta_{K}(\Lambda_K)+\delta_S(\eta)~,
\end{align}
where $\xi^{\mu}=2\bar{\epsilon}_{2}^{i}\gamma^{\mu}\epsilon_{1i}+\text{h.c.}$ while
$\varepsilon$, $\eta$, and $\Lambda_K$ depend on the auxiliary fields as
\begin{align}\label{modified_structure}
\varepsilon_{ab} &= \varepsilon^{ij}\bar{\epsilon}_{1i}\epsilon_{2j}T^{-}_{ab} + \text{h.c.}~,\nonumber \\
\Lambda^a_{K}&=\epsilon_1^i\epsilon_2^jD_bT^{ba}_{ij}-\bar{\epsilon}_2^i\gamma^a\epsilon_{1i}D-2\bar{\epsilon}_2^i\gamma^b\epsilon_{1i}T_{bc}{}^{jk}T^{ac}{}_{jk}+\thc~,\nonumber\\
\eta^i&=6\bar{\epsilon}^i_{[1}\epsilon_{2]}^j\chi_j-\bar{\epsilon}_{[2}^j\gamma^a\epsilon_{1]j}\gamma_a\chi^i~.
\end{align}
The shifts in the connections (\ref{changes}) have induced shifts in $\L_K$ and $\eta^i$ relative
to the standard constraints. This is simply because when one shifts connections within
the covariant diffeomorphism $\delta^{(cov)}(\xi)$, without redefining $\delta_Q$ itself,
the shifts must appear as explicit new terms in the above commutator. We emphasize
that because neither $\delta_Q$ nor $\delta_S$ have been redefined, their commutator
remains unchanged,
\begin{align}
\left[\delta_S(\eta),\delta_Q(\epsilon)\right]&=\delta_{M}(\bar{\eta}_{i}\gamma^{ab}\epsilon^{i}+\text{h.c.})+\delta_{D}(\bar{\eta}^{i}\epsilon_{i}+\text{h.c.})\nonumber \\
&\quad
+\delta_A(i\bar{\eta}_{i}\epsilon^{i}+\text{h.c.})
+\delta_{V}(-2\bar{\eta}^{i}\epsilon_{j}-(\text{h.c. ; traceless}))~.
\end{align}

\section{The vector multiplet and the multiplet of its equations of motion}
Now let us introduce an abelian vector multiplet. This is an off-shell representation of the superconformal algebra, transforming under Q and S-supersymmetries and K-boosts as
\begin{align}
\label{vectortransfmod}
\delta X &= \bar{\epsilon}^{i}\Omega_{i}\nonumber \\
\delta\Omega_{i}&=2\Dslash X \epsilon_{i}+\frac{1}{2}\epsilon_{ij}\gamma\cdot \hat{\mathcal{F}}\epsilon^{j}+Y_{ij}\epsilon^{j}+2X\eta_{i} \nonumber \\
\delta W_{\mu}&=\varepsilon^{ij}\bar{\epsilon}_{i}\gamma_{\mu}\Omega_{j}+2\varepsilon_{ij}\bar{X}\bar{\epsilon}^{i}\psi_{\mu}^{j}+\thc\nonumber \\
\delta Y_{ij}&=2\bar{\epsilon}_{(i}\Gamma_{j)}+2\varepsilon_{ik}\varepsilon_{jl}\bar{\epsilon}^{(k}\Gamma^{l)}
\end{align}
with
\begin{align}\label{def}
\Gamma_{j}&=\Dslash\Omega_{j}-2X\chi_{j}~, \qquad
\hat{\mathcal{F}}_{\mu\nu}=\mathcal{F}_{\mu\nu}-\frac{1}{4}\varepsilon_{ij}\bar{X}T_{\mu\nu}{}^{ij}-\frac{1}{4}\varepsilon^{ij}XT_{\mu\nu ij}\;.
\end{align}
We have written these transformation rules already using the new composite connections
associated with the unconventional constraints \eqref{unconven_constraints}.

In the above equation, $\mathcal{F}_{\mu\nu}$ is the superconformally covariant field strength associated with the gauge field $W_\mu$ and is given as
\begin{align*}
\mathcal{F}_{\mu\nu}&=2\partial_{[\mu}W_{\nu]}-\varepsilon_{ij}\bar{\psi}_{[\mu}{}^{i}\Big(\gamma_{\nu]}\Omega^{j}+\psi_{\nu]}{}^j \bar X\Big)-\varepsilon^{ij}\bar{\psi}_{[\mu i}\Big(\gamma_{\nu]}\Omega_{j}+\psi_{\nu]j} X\Big)\;.
\end{align*}
The supercovariant field strength satisfies the following Bianchi identity
\begin{align}\label{Bianchi}
D^{b}\left(\mathcal{F}_{ab}^{+}-\mathcal{F}_{ab}^{-}\right)=0
\end{align}
or in terms of $\hat{\mathcal{F}}$, 
\begin{align*}
D^{b}\left[\hat{\mathcal{F}}_{ab}^{+}-\hat{\mathcal{F}}_{ab}^{-}+\frac{1}{4}\varepsilon^{ij}X T_{abij}-\frac{1}{4}\varepsilon_{ij}\bar{X} T_{ab}{}^{ij}\right]=0\;.
\end{align*}
This multiplet contains $8+8$ degrees of freedom off-shell, and $4+4$ on-shell (for the bosons, two from the complex scalar plus two from the on-shell gauge field).

We add here that because the vector multiplet contains a gauge field $W_{\mu}$, the
algebra of supersymmetry transformations \eqref{algebra} will in general contain also a local
gauge transformation $\delta_{gauge}(\lambda)$ with field-dependent parameter $\lambda$.
Computing $\left[\delta_Q,\delta_Q\right]$ on $W_\mu$, one finds
\begin{align}\label{theta}
\lambda=4\varepsilon^{ij}\bar{\epsilon}_{2i}\epsilon_{1j}X+\text{h.c.}
\end{align}

At this point we want to obtain the constraints which will allow us to trade the auxiliary fields of the standard Weyl multiplet for the fields  of the vector  multiplet. As  was explained  in \cite{Bergshoeff:2001hc} there are two  ways of achieving this: 
one is to use the superconformally invariant action of the vector multiplet coupled to the standard Weyl multiplet and eliminate the auxiliary fields via the equations of motion. The second procedure, as outlined in \cite{Bergshoeff:2001hc}, is more suitable for getting the equations of motion directly without referring to the action. In this procedure, we obtain the multiplet of equations of motion by starting from the equation of motion with the lowest Weyl weight, which is also S-invariant. The other components of the equation of motion multiplet are obtained by successive application of Q-supersymmetry.

In our case, the starting point of  the analysis is the algebraic equation of motion of $Y_{ij}$ which has the lowest Weyl weight $w=2$. The only covariant combination of fields of Weyl weight 2 that we can write down, which is also S-invariant, is $Y_{ij}=0$. Successive applications of Q-supersymmetry give
\begin{align}\label{eom}
\delta Y_{ij}&=2\bar{\epsilon}_{(i}\Gamma_{j)}+2\varepsilon_{ik}\varepsilon_{jl}\bar{\epsilon}^{(k}\Gamma^{l)}\\
\delta \Gamma_{j}&=2 \Big[\bar{\chi}^{k}\Omega_{k}+D^2 X - XD+\frac{1}{8}\hat{\mathcal{F}}\cdot T^{-}\Big]\epsilon_{j}+2\varepsilon_{jk}D_a \Big[\hat{\mathcal{F}}^{-ab}-\frac{1}{4}XT^{+ab}\Big]\gamma_{b}\epsilon^{k}+\Dslash Y_{jk}\epsilon^{k}\;. \nonumber
\end{align}
The multiplet of equations of motion is then $\{Y_{ij},\G_i, \cC, \cJ^a\}$, with the final constraints reading:
\begin{align}\label{eom1}
Y_{ij}&=0\nonumber \\
\Gamma_{i}&= \Dslash\Omega_{i}-2X\chi_{i}=0\nonumber \\
\mathcal{C}&= \bar{\chi}^{i}\Omega_{i}+D^2 X - XD+\frac{1}{8}\hat{\mathcal{F}}\cdot T^{-}=0 \nonumber \\
\mathcal{J}^{a}&=-D_b \Big(\hat{\mathcal{F}}^{-ab}-\frac{1}{4}XT^{+ab}\Big)=0~.
\end{align}
Here we have defined
\begin{align}\label{}
T^{+}_{ab}=\varepsilon^{ij}T_{abij}, \quad T^{-}_{ab}=\varepsilon_{ij}T_{ab}{}^{ij}~.
\end{align}
The equation $\mathcal{J}^{b}=0$ is a complex equation. Its imaginary part is automatically satisfied by the Bianchi identity \refb{Bianchi}, but its real part gives a non-trivial equation,
\begin{align}\label{Heqn}
\cJ^b + \bar \cJ^b = D_a \left[{\mathcal{F}}^{ab}-\frac12 XT^{+ab}-\frac12 \bar{X}T^{-ab}\right]=0~.
\end{align}
This can be interpreted as the Bianchi identity of a different vector gauge field $\tilde{W}_{\mu}$. In particular, we can write the equation \refb{Heqn} as the Bianchi identity 
\begin{align}\label{BBianchi}
D_{[a}\mathcal{G}_{bc]}=0
\end{align}
with $\mathcal{G}_{\mu\nu}$ given by
\begin{align}\label{Gdef}
\frac{i}{2}\varepsilon_{abcd}\mathcal{G}^{cd}&=\mathcal{F}_{ab}-\frac12 XT^{+ab}-\frac12 \bar{X}T^{-ab}~,
\end{align}
or equivalently
\begin{align}\label{gdef1}
i\left(\mathcal{G}^{+ab}-\mathcal{G}^{-ab}\right)&=\mathcal{F}_{ab}- \frac12 XT^{+ab}-\frac12 \bar{X}T^{-ab}~.
\end{align}
Here, $\mathcal{G}_{\mu\nu}$ should be interpreted as the supercovariant field strength of a new vector gauge field $\tilde{W}_{\mu}$. Using the above equation one can solve for $T_{ab}^\pm$ in terms of the field strength of $W_{\mu}$ and $\tilde{W}_{\mu}$ as follows
\begin{align}\label{T}
T^{+ab}&=2\,X^{-1}\left(\mathcal{F}^{+ab}-i\mathcal{G}^{+ab}\right), \quad T^{-ab}=2\,\bar{X}^{-1}\left(\mathcal{F}^{-ab}+i\mathcal{G}^{-ab}\right)~.
\end{align}
The supersymmetry transformation of $\mathcal{G}_{ab}$ can be obtained from the definition (\ref{Gdef}) using the transformations of $T_{ab}^\pm$ and $\mathcal{F}_{ab}$. Then one can work out the transformation of $\tilde{W}_{\mu}$ from the knowledge of the supersymmetry transformation of $\mathcal{G}_{\mu\nu}$, the Bianchi identity ($\ref{BBianchi}$) and supersymmetry algebra (\ref{algebra}). The result is
\begin{align}\label{Btransf}
\delta \tilde{W}_{\mu}&=i\varepsilon^{ij}\bar{\epsilon}_{i}\gamma_{\mu}\Omega_{j}-2i\varepsilon_{ij}\bar{X}\bar{\epsilon}^{i}\psi_{\mu}^{j}+\thc
\end{align}
From the above transformation law, one can work out the form of the superconformally invariant field strength $\mathcal{G}_{\mu\nu}$,
\begin{align}\label{Bfieldstrength}
\mathcal{G}_{\mu\nu}&=2\partial_{[\mu}\tilde{W}_{\nu]}+i\varepsilon_{ij}\bar{\psi}_{[\mu}{}^{i}\Big(\gamma_{\nu]}\Omega^{j}+\psi_{\nu]}{}^j \bar X\Big)-i\varepsilon^{ij}\bar{\psi}_{[\mu i}\Big(\gamma_{\nu]}\Omega_{j}+\psi_{\nu]j} X\Big)
\end{align}
The $\left[\delta_Q,\delta_Q\right]$ commutator picks up a field dependent gauge transformation $\delta_{gauge}(\tilde{\lambda})$ when it acts on $\tilde{W}_{\mu}$, with 
\begin{align}\label{tildetheta}
\tilde{\lambda}=-4i \,\varepsilon^{ij}\bar{\epsilon}_{2i}\epsilon_{1j}X+\text{h.c.}
\end{align} 

Next, we use the equation $\Gamma_{j}=0$, to eliminate $\chi_{j}$ as
\begin{align}\label{chi}
\chi_{j}&=\frac{1}{2}X^{-1}\Dslash\Omega_{j}
\end{align}

Finally, using the real and imaginary parts of the $\mathcal{C}=0$ equation of motion,
we can eliminate $D$ and find a constraint on the complex scalar $X$,
\begin{align}\label{D}
&D=\frac{1}{2}|X|^{-2}\left(XD^2 \bar{X}+\frac{1}{2}\bar{\Omega}^{k}\Dslash \Omega_{k}+\frac{1}{4}\mathcal{F}\cdot\mathcal{F}^{+}+\frac{1}{4}\mathcal{G}\cdot\mathcal{G}^{+}+\thc\right)~, \\
\label{constraint}
&XD^2 \bar{X}+\frac{1}{2}\bar{\Omega}^{k}\Dslash \Omega_{k}+\frac{1}{4}\mathcal{F}\cdot\mathcal{F}^{+}+\frac{1}{4}\mathcal{G}\cdot\mathcal{G}^{+}-\thc=0\;.
\end{align}

In the next subsection we will  analyze this last constraint in detail, and show that it can be used to trade the U(1)$_R$ connection for a two form gauge field.

\subsection{Analysis  of the constraint on the scalar fields}
The constraint \eqref{constraint} can be easily re-written in the form
\begin{equation}
D^a(-X\bar X\,D_a \log \frac{X}{\bar X}+\tfrac12\,\bar\Omega^k\gamma_a\Omega_k)=-\frac{1}{8}\cF\cdot\tilde\cF-\frac{1}{8}\cG\cdot\tilde\cG
\end{equation}
Let us denote the term in the parentheses as
\begin{equation}\label{Hdef}
X\bar X\,D_a \log \frac{X}{\bar X}-\tfrac12\,\bar\Omega^k\gamma_a\Omega_k=\tfrac1{3!}\varepsilon_{abcd}\,\cH^{bcd}~.
\end{equation}
With the above definition, the constraint (\ref{constraint}) becomes a Bianchi identity for  the three-form field strength $\cH$ as shown below.
\begin{equation}
\label{eq:Bianchi_identity_new}
D_{[a}\cH_{bcd]}=\tfrac38\cF_{[ab}\cF_{cd]}+\tfrac38\cG_{[ab}\cG_{cd]}~.
\end{equation}
Now expanding out the covariant derivative in \eqref{Hdef} gives
\begin{equation}
\label{eq:u1gaugefield}
2{\rm i}X\bar X\,A_a+X\bar X\,\partial_a \log \frac{X}{\bar X}-\tfrac12\,\bar X\,\bar\psi^i_{a}\Omega_i +\tfrac12\,X\,\bar\psi_{ai}\Omega^i-\tfrac12\,\bar\Omega^k\gamma_a\Omega_k=\tfrac1{3!}\varepsilon_{abcd}\,\cH^{bcd}~.
\end{equation}
We can then interpret \eqref{eq:u1gaugefield} as a constraint that determines the U(1)$_R$ symmetry gauge field in terms of the other fields.
Because $A_\mu$ is a gauge field, this is only possible because it combines with the phase
of the complex scalar $X$; the ``Higgsed'' U(1)$_R$ gauge field is what is determined
by the above combination.

We can further use (\ref{Hdef}) to obtain the U(1)$_R$ symmetry field strength in terms of the three-form field strength and other covariant objects as
\begin{align}\label{RA}
R(A)_{ab}&=\frac{i}{4}|X|^{-2}\varepsilon_{abfc}D_{d} \cH^{fcd}-\frac{i}{4X}\bar{R}(Q)_{ab}^{i}\Omega_{i}+\frac{i}{4\bar{X}}\bar{R}(Q)_{abi}\Omega^{i}-2i|X|^{-2}D_{[a}XD_{b]}\bar{X}\nonumber \\
&\quad-\frac{i}{2}|X|^{-2}D_{[a}\left(\Omega^{k}\gamma_{b]}\Omega_{k}\right)\;.
\end{align}
The dual of this appears in the $R(M)$ constraint in (\ref{unconven_constraints}).

Now we must solve the Bianchi identity \eqref{eq:Bianchi_identity_new} for a
two-form field $B_{\mu\nu}$ just as we solved  \eqref{BBianchi} for $\tilde W_\mu$.
For the non-supersymmetric case, the solution to the Bianchi identity would be
\begin{align}\label{E:DefH}
H_{\mu\nu\rho}&:=3\partial_{[\mu} B_{\nu\rho]}+\tfrac34 W_{[\mu}F_{\nu\rho]}+\tfrac34 \tilde{W}_{[\mu}G_{\nu\rho]}\;,
\end{align}
where $F_{\mu\nu}=2\partial_{[\mu}W_{\nu]}$ and $G_{\mu\nu}=2\partial_{[\mu}\tilde{W}_{\nu]}$ are the standard field strengths of the gauge fields $W_{\mu}$ and $\tilde{W}_{\mu}$ respectively. We expect $\cH$ to differ from $H$ by supercovariantizations involving the gravitino which we find below.

The two-form $B_{\mu\nu}$ should transform as the exterior derivative of a one-form
as well as under the zero-form gauge transformations corresponding to $W_{\mu}$ and $\tilde{W}_{\mu}$.
In addition, it should transform under supersymmetry.
From \eqref{Hdef}, one can find that the three form field strength $\cH_{abc}$
transforms under supersymmetry as
\begin{align}\label{susy_transf_H}
\delta_{\text{Q}}\Big(\frac{1}{6}\varepsilon_{abcd}\cH^{bcd}\Big)&=-\bar{\epsilon}^i\gamma_{ab}D^b(\Omega_i\bar{X})+\frac{1}{16}\varepsilon_{ij}\bar{\epsilon}^i\gamma_a\gamma\cdot(\mathcal{F}^+-i\mathcal{G}^+)\Omega^j\nonumber\\
&\quad+\frac{1}{8}\varepsilon_{ij}\bar{\epsilon}^i\gamma\cdot(\mathcal{F}^--i\mathcal{G}^-)\gamma_a\Omega^j-\thc \nonumber\\
\delta_{\text{S}}\Big(\frac{1}{6}\varepsilon_{abcd}\cH^{bcd}\Big)&=-\frac{3}{2}\bar{X}\bar{\eta}^{i}\gamma_{a}\Omega_{i}-\thc
\end{align}
Of course, $\cH$ is inert under the one-form transformation of $B_{\mu\nu}$ as well
as the zero-form transformations of $W_\mu$ and $\tilde{W}_\mu$. To obtain $\delta B_{\mu\nu}$ we
require
\begin{enumerate}
\item
the Bianchi identity of the three form field strength as given in (\ref{eq:Bianchi_identity_new}),
\item
the supersymmetry transformation of the three form field strength as given in (\ref{susy_transf_H}),
\item
the closure of the supersymmetry algebra, both $\left[\delta_Q,\delta_Q\right]$ as well as $\left[\delta_S,\delta_Q\right]$.
\end{enumerate}
After incorporating the above operations, we obtain
\begin{align}\label{varC}
\delta B_{\mu\nu}&=\tfrac12 W_{[\mu} \delta_Q W_{\nu]}+\tfrac12 \tilde{W}_{[\mu} \delta_Q \tilde{W}_{\nu]}+\bar X \bar\epsilon^i\gamma_{\mu\nu}\Omega_i+ X\bar\epsilon_i\gamma_{\mu\nu} \Omega^i+2\,X\bar X\bar\epsilon^i\gamma_{[\mu}\psi_{\nu]\,i}+2\,X\bar X \bar\epsilon_i\gamma_{[\mu}\psi_{\nu]}^i\nonumber \\
&\quad+2\partial_{[\mu}\Lambda_{\nu]}-\frac{\lambda}{4}{F}_{\mu\nu}-\frac{\tilde{\lambda}}{4}{G}_{\mu\nu}\;.
\end{align}
The parameters $\lambda$ and $\tilde{\lambda}$ are the gauge transformation parameters associated with the gauge fields $W_{\mu}$ and $\tilde{W}_{\mu}$. The above analysis also shows that $B_{\mu\nu}$ is S-invariant. From \eqref{Hdef}, it also follows that $B_{\mu\nu}$ has zero Weyl and chiral weights. From the above transformation of $B_{\mu\nu}$, one can obtain the explicit form of the fully supercovariant  three form field strength $\cH_{\mu\nu\rho}$ as 
\begin{align}\label{3formfieldstrength}
\cH_{\mu\nu\rho}&:= H_{\mu\nu\rho}-\tfrac32\bar X\bar\psi_{[\mu}^i\gamma_{\nu\rho]}\Omega_i-\tfrac32 X \bar\psi_{[\mu\,i}\gamma_{\nu\rho]}\Omega^i-3 X\bar X \bar\psi_{[\mu}^i\gamma_\nu\psi_{\rho]}{}_i~,
\end{align}
with $H_{\mu\nu\rho}$ given by \eqref{E:DefH}.
The commutator $\left[\delta_Q,\delta_Q\right]$ picks up a field dependent vector gauge transformation $\delta_{V}(\Lambda_{\mu})$ when it acts on $B_{\mu\nu}$, with
\begin{align}\label{vector-gauge-parameter}
\Lambda_\mu&=\varepsilon_{ij}\bar{\epsilon}_1^i\epsilon_2^j\bar{X}W_\mu-i\varepsilon_{ij}\bar{\epsilon}_1^i\epsilon_2^j\bar{X} \tilde W_\mu + 2X\bar{X}\bar{\epsilon}_2^i\gamma_\mu\epsilon_{1i}+\thc
\end{align}

\section{The $N=2$ dilaton Weyl multiplet in components}
Let us summarize our result for the $N=2$ dilaton Weyl multiplet now.
It possesses 24+24 degrees of freedom. 
The independent bosonic fields are as follows: the vierbein $e_{\mu}{}^{a}$, the dilatation gauge field $b_{\mu}$, the $\gSU(2)$-gauge field $\mathcal{V}_{\mu}{}^{i}{}_{j}$, a complex scalar $X$, two abelian gauge fields $W_{\mu}$ and $\tilde{W}_{\mu}$, and a two-form tensor gauge field $B_{\mu\nu}$. 
The fermionic field content is as follows: the gravitino, whose left and right chiral components are denoted by the SU(2) doublets $\psi_{\mu}{}_{i}$ and $\psi_{\mu}{}^{i}$ and
another SU(2) doublet of spinors whose left and right chiral components are denoted by $\Omega^{i}$ and $\Omega_{i}$.

Apart from the generators of the $\gSU(2,2|2)$ superconformal algebra, this multiplet also has one vector gauge symmetry corresponding to transformation of $B_{\mu\nu}$ and two U(1) gauge symmetries corresponding to the gauge fields $W_{\mu}$ and $\tilde{W}_{\mu}$. The latter two will be denoted U(1)$_W$ and U(1)$_{\tilde{W}}$ respectively. 

In what follows, we give how the independent fields transform under all the superconformal transformations and the other gauge transformations present in the dilaton Weyl multiplet: Q-susy, S-susy, K-boosts, local Lorentz transformation, $\gSU(2)\times \gU(1)$ R-symmetry, dilatation, U(1)$_W$, U(1)$_{\tilde{W}}$ and vector-gauge transformation parameterized by $\epsilon^{i}$, $\eta^{i}$, $\Lambda_{K}^{a}$, $\Lambda_{M}^{ab}$, $(\Lambda^{i}{}_{j},\Lambda_A)$, $\Lambda_D$, $\lambda$, $\tilde{\lambda}$ and $\Lambda_{\mu}$ respectively:
\begin{align}\label{dilatontransf}
\delta e_{\mu}{}^{a}&=\bar{\epsilon}^{i}\gamma^{a}\psi_{\mu i}+\thc-\Lambda_D e_{\mu}{}^{a}+\Lambda_{M}^{ab}e_{\mu b} \nonumber \\
\delta\psi_{\mu}{}^{i}&=2\mathcal{D}_{\mu}\epsilon^{i}-\frac{1}{16}\varepsilon^{ij}\bar{X}^{-1}\gamma\cdot \left(\mathcal{F}^{-}+i\mathcal{G}^{-}\right)\gamma_{\mu}\epsilon_{j}-\gamma_{\mu}\eta^{i}-\frac{1}{2}\Lambda_D\psi_{\mu}{}^{i}-\frac{i}{2}\Lambda_A\psi_{\mu}{}^{i}+\Lambda^{i}{}_{j}\psi_{\mu}{}^{j}
\nonumber \\
&\quad +\frac{1}{4}\Lambda_{M}^{ab}\gamma_{ab}\psi_{\mu}{}^{i}\nonumber \\
\delta b_{\mu}&=\frac{1}{2}\bar{\epsilon}^{i}\phi_{\mu i}-\frac{1}{4}X^{-1}\bar{\epsilon}^{i}\gamma_{\mu}\Dslash\Omega_{i}-\frac{1}{2}\bar{\eta}^{i}\psi_{\mu i}+\thc+\Lambda_{K}^{a}e_{\mu a} +\partial_{\mu}\Lambda_D\nonumber \\
\delta \mathcal{V}_{\mu}{}^{i}{}_{j}&=2\bar{\epsilon}_{j}\phi_{\mu}^{i}-\bar{X}^{-1}\bar{\epsilon}_{j}\gamma_{\mu}\Dslash\Omega^{i}+2\bar{\eta}_{j}\psi_{\mu}^{i}-(\thc; ~\text{traceless})-2\partial_{\mu}\Lambda^{i}{}_{j}+\Lambda^{i}{}_{k}\mathcal{V}_{\mu}{}^{k}{}_{j}-\Lambda^{k}{}_{j}\mathcal{V}_{\mu}{}^{i}{}_{k}\nonumber \\
\delta X &= \bar{\epsilon}^{i}\Omega_{i}+\left(\Lambda_D -i\Lambda_A\right)X \nonumber \\
\delta\Omega_{i}&=2\Dslash X \epsilon_{i}+\frac{1}{4}\epsilon_{ij}\gamma\cdot {\mathcal{F}}\epsilon^{j}-\frac{i}{4}\epsilon_{ij}\gamma\cdot \mathcal{G}^{-}\epsilon^{j}+2X\eta_{i} +\left(\frac{3}{2}\Lambda_D -\frac{i}{2}\Lambda_A\right)\Omega_{i}-\Lambda^{j}{}_{i}\Omega_{j}\nonumber \\
&\quad +\frac{1}{4}\Lambda_{M}^{ab}\gamma_{ab}\Omega_{i}\nonumber \\
\delta W_{\mu}&=\varepsilon^{ij}\bar{\epsilon}_{i}\gamma_{\mu}\Omega_{j}+2\varepsilon_{ij}\bar{X}\bar{\epsilon}^{i}\psi_{\mu}^{j}+\thc+\partial_{\mu}\lambda\nonumber \\
\delta \tilde{W}_{\mu}&=i\varepsilon^{ij}\bar{\epsilon}_{i}\gamma_{\mu}\Omega_{j}-2i\varepsilon_{ij}\bar{X}\bar{\epsilon}^{i}\psi_{\mu}^{j}+\thc+\partial_{\mu}\tilde{\lambda}
\nonumber\\
\delta B_{\mu\nu}&=\tfrac12 W_{[\mu} \delta_Q W_{\nu]}+\tfrac12 \tilde{W}_{[\mu} \delta_Q \tilde{W}_{\nu]}+\bar X \bar\epsilon^i\gamma_{\mu\nu}\Omega_i+ X\bar\epsilon_i\gamma_{\mu\nu} \Omega^i+2\,X\bar X\bar\epsilon^i\gamma_{[\mu}\psi_{\nu]\,i}+2\,X\bar X \bar\epsilon_i\gamma_{[\mu}\psi_{\nu]}^i \nonumber \\
&\quad +2\partial_{[\mu}\Lambda_{\nu]}-\frac{\lambda}{4}{F}_{\mu\nu}-\frac{\tilde{\lambda}}{4}{G}_{\mu\nu}
\end{align}
where
\begin{align}\label{delmuepsilon}
\mathcal{D}_{\mu}\epsilon^{i} = 
	\partial_{\mu}\epsilon^{i}
	-\frac{1}{4}\omega_{\mu}{}^{ab}\gamma_{ab}\epsilon^{i}
	+\frac{1}{2}\left(b_{\mu}+iA_{\mu}\right)\epsilon^{i}
	+\frac{1}{2}\mathcal{V}_{\mu}{}^{i}{}_{j}\epsilon^{j}~.
\end{align}
The gauge fields $\omega_{\mu}{}^{ab}$, $f_{\mu}^{a}$, $\phi_{\mu}^{i}$ and $A_{\mu}$ corresponding to local Lorentz transformation, K-boosts, S-supersymmetry and U(1)$_R$ symmetry are composite gauge fields which are determined from the constraints \eqref{unconven_constraints} and \eqref{eq:u1gaugefield}. The U(1)$_R$ field strength that appears in the constraint (\ref{unconven_constraints}) is given in terms of the three-form field strength and other covariant objects as shown in (\ref{RA}). The $\left[\delta_Q,\delta_Q\right]$ commutator acts on this multiplet as
\begin{align}\label{dilatonalgebra}
\left[\delta_Q(\epsilon_1),\delta_Q(\epsilon_2)\right]=\delta^{(cov)}(\xi)+\delta_{M}(\varepsilon)+\delta_{K}(\Lambda_K)+\delta_S(\eta)+\delta_{W}(\lambda)+\delta_{\tilde{W}}(\tilde{\lambda})+\delta_{V}(\Lambda_{\mu})
\end{align}
where the parameters $\varepsilon$, $\Lambda_K$, and $\eta$ are as given in (\ref{modified_structure}), $\lambda$ is as given in (\ref{theta}), $\tilde{\lambda}$ is as given in (\ref{tildetheta}) and $\Lambda_{\mu}$ is as given in (\ref{vector-gauge-parameter}). Here $\delta_W$, $\delta_{\tilde{W}}$ and $\delta_V$ corresponds to transformation under U(1)$_W$, U(1)$_{\tilde{W}}$ and vector-gauge transformation respectively.

A few comments about this multiplet are in order. As with the dilaton Weyl multiplet in five dimensions, it possesses a scalar field that can be used to gauge-fix the Weyl symmetry. However, here that field appears in a complex scalar $X$ together with a phase that can be used to gauge-fix the $\gU(1)_R$ symmetry. The Weyl fermions $\Omega_i$ can similarly be used to gauge-fix S-supersymmetry. As with the standard Weyl multiplet, $b_\mu$ may be eliminated by a choice of K-gauge.

\section{The $N=2$ dilaton Weyl multiplet in superspace}

The 4d $N=2$ dilaton Weyl multiplet was already constructed in superspace
by M\"uller in \cite{Muller:1986ku}, but with the addition of a compensating
tensor multiplet that accomplishes the breaking of SU(2)$_R$ to SO(2)$_R$. The result is
a minimal $32+32$ supergravity multiplet. The further breakdown of the $R$-symmetry group,
along with the additional component fields, seems to have prevented its appreciation.
Here we give a modified version of his superspace construction, with the additional
tensor multiplet removed, that leads to the dilaton Weyl multiplet.
Aside from notational changes, we also employ $N=2$
conformal superspace \cite{Butter:2011sr} (we follow the conventions in \cite{Butter:2013lta}),
whose component formulation can be exactly identified with $N=2$ conformal
supergravity. Note that the 5d dilaton Weyl multiplet has also been constructed
in superspace, see \cite{Kuzenko:2008wr} and \cite{Butter:2014xxa}.

Recall in superspace that a vector multiplet is described by a closed two-form $\cF = \rd \cW$
whose tangent space components are given by
\begin{align}
\cF_{\alpha i\, \beta j} &= -4 \eps_{\alpha \beta} \veps_{ij} \bar \cX~, \qquad\quad
\cF^{\dalpha i\, \dbeta j} = -4 \eps^{\dalpha \dbeta} \veps^{ij} \cX~, \eol
\cF_{\alpha i\, b} &= (\gamma_{b})_{\alpha \dalpha} \veps_{i j} \bar\nabla^{\dalpha j} \bar \cX~, \qquad
\cF^{\dalpha i}{}_{b} = (\gamma_{b})^{\dalpha \alpha} \veps^{i j} \nabla_{\alpha j} \cX~, \eol
\cF_{a b} &= -\frac{1}{8} (\gamma_{ab})^{\a\b} \nabla_{\a\b} \cX
	- \frac{1}{8} (\gamma_{ab})_{\ad\bd} \bar\nabla^{\ad\bd} \bar \cX
	+ W_{ab}^+ X + W_{a b}^- \bar \cX~.
\end{align}
The superspace Bianchi identity $\rd \cF = 0$ imposes that the superfield $\cX$ should be
a reduced chiral, meaning that it obeys the two constraints
\begin{align}
\bar\nabla^{\ad i} \cX = 0~, \qquad
\nabla_{i j} \cX = \veps_{i k} \veps_{j l} \bar\nabla^{kl} \bar \cX~.
\end{align}
If the superfield obeys in addition the on-shell constraint $\nabla_{ij} \cX = 0$,
one can introduce a second closed two-form $\cG$ with $\cX$ replaced by $i \cX$,
\begin{align}
\cG_{\alpha i\, \beta j} &= 4 i \eps_{\alpha \beta} \veps_{ij} \bar \cX~, \qquad\qquad\,\,\,\,
\cG^{\dalpha i\, \dbeta j} = -4 i \eps^{\dalpha \dbeta} \veps^{ij} \cX~, \eol
\cG_{\alpha i\, b} &= - i (\gamma_{b})_{\alpha \dalpha} \veps_{i j} \bar\nabla^{\dalpha j} \bar \cX~, \qquad
\cG^{\dalpha i}{}_{b} = i (\gamma_{b})^{\dalpha \alpha} \veps^{i j} \nabla_{\alpha j} \cX~, \eol
\cG_{a b} &= -\frac{i}{8} (\gamma_{ab})^{\a\b} \nabla_{\a\b} \cX
	+ \frac{i}{8} (\gamma_{ab})_{\ad\bd} \bar\nabla^{\ad\bd} \bar \cX
	+ i W_{ab}^+ \cX - i W_{a b}^- \bar \cX~.
\end{align}
The closure of $\cG$ implies that (at least locally) $\cG = \rd \tilde{\cW}$ for a new one-form $\tilde{\cW}$.

Following \cite{Muller:1986ku}, $\cF$ and $\cG$ can now be combined into a complex super
two-form $\cF + i \cG$,
\begin{align}
(\cF+i \cG)_{\alpha i\, \beta j} &= -8 \eps_{\alpha \beta} \veps_{ij} \bar \cX~, \qquad
(\cF+i \cG)^{\dalpha i\, \dbeta j} = 0~, \eol
(\cF+i \cG)_{\alpha i\, b} &= 2 (\gamma_{b})_{\alpha \dalpha} \bar\nabla^\dalpha \bar \cX~, \qquad
(\cF+i \cG)^{\dalpha i}{}_{b} = 0~, \eol
(\cF+i \cG)_{a b} &=
	- \frac{1}{4} (\gamma_{ab})_{\ad\bd} \bar\nabla^{\ad\bd} \cX
	+ 2 \bar \cX W_{a b}^-~.
\end{align}
Encoded in the anti-self-dual part of the last equation is a constraint on the super-Weyl
tensor $W_{ab}^-$,
\begin{align}
\cF_{ab}^- + i \cG_{ab}^- &= 2 \bar \cX W_{a b}^-~.
\end{align}
This is the superspace version of the component equation \eqref{T}, with
$T_{ab}^- = 4 \,W_{ab}^-\vert_{\q=0}$. The component constraints on $\chi^i$ and $D$ follow
by applying superspace spinor derivatives.

At the component level, the Higgsed $\gU(1)_R$ connection has been exchanged for
a two-form. This two-form can be identified directly in superspace by constructing its three-form
field strength,\footnote{In this section,
we employ superspace conventions for differential forms, see e.g. \cite{Wess:1992cp}.}
\begin{align}\label{E:DefSuperH}
\cH &= \rd \cB + \frac{1}{4} \cW \wedge \cF + \frac{1}{4} \tilde{\cW} \wedge \cG~, \qquad
\rd \cH = \frac{1}{4} \cF \wedge \cF + \frac{1}{4} \cG \wedge \cG~.
\end{align}
$\cH$ is gauge invariant which implies that $\cB$ transforms as
\begin{align}\label{eq:BGauge}
\delta \cB = \rd \L - \frac{1}{4} \lambda \,\cF - \frac{1}{4} \tilde{\lambda} \,\cG~,
\end{align}
where $\L$ is the one-form gauge transformation parameter and $\lambda$ and $\tilde{\lambda}$
are the zero-form gauge transformation parameters of $\cW$ and $\tilde{\cW}$.
(By redefining the $\Lambda$ transformation, one could write the $\lambda$ gauge transformation as $\rd \lambda \wedge \cW$ and similarly for $\tilde{\lambda}$.)
The dimension-3/2 and dimension-2 tangent space components of $\cH$ are given by
\begin{subequations}
\begin{align}
\cH_{\ul {\gamma \beta \alpha}} = 0~, \qquad
\cH_{\gamma k\, \beta j, a} = \cH^{\dgamma k\, \dbeta j}{}_{a} = 0~, \qquad
\cH_{\gamma k}{}^{\dbeta j}{}_a = - 2\, \delta_k^j (\gamma_a)_\gamma{}^\dbeta \cX \bar \cX~.
\end{align}
The remaining pieces are given by
\begin{align}
\cH_{a b \alpha i} &= (\gamma_{a b})_\alpha{}^\beta \nabla_{\beta i} (\cX \bar \cX)~, \qquad
\cH_{a b}{}^{\dalpha i} = (\gamma_{a b})^\dalpha{}_\dbeta \bar\nabla^{\dbeta i} (\cX \bar \cX)~, \\
\cH_{a b c} &= 
	\veps_{a b c d} \Big(
	(\cX \nabla^d \bar \cX - \bar \cX \nabla^d \cX)
	- \frac{1}{2} (\gamma^d)_{\alpha \dalpha} \nabla^\alpha_k \cX \bar\nabla^{\dalpha k} \bar \cX
	\Big)~.\label{E:SuperH.d}
\end{align}
\end{subequations}

The connection between superspace and component results is straightforward.
The component three-form $H_{\mu\nu\rho}$ is identified by projecting the super-three-form
$\cH$ to $\theta=\rd\theta=0$, i.e.  $H_{\mu\nu\rho} = \cH_{\mu\nu\rho}\vert_{\q=0}$, so
that \eqref{E:DefSuperH} reduces to \eqref{E:DefH}.
The supercovariant field strength $\cH_{abc}$ \eqref{3formfieldstrength} corresponds to
the $\theta=0$ projection of \eqref{E:SuperH.d}.
The $Q$-supersymmetry transformation of $B$ descends from a covariant Lie derivative of $\cB$ in superspace, that is, the usual Lie derivative with parameter 
$\xi^A = (0, \eps^{\a i}, \eps_{\dalpha i})$
accompanied by the gauge transformations
\begin{align}
\lambda = - \imath_\xi \cW ~, \qquad \tilde{\lambda} = -\imath_\xi \tilde{\cW}~, \qquad \L= -\imath_\xi \cB~.
\end{align}
This leads to
\begin{align}
\delta^{\rm cov}_\xi \cB &= 
	\rd \imath_\xi \cB + \imath_\xi \rd \cB
	+ \rd \L - \frac{1}{4} \lambda \wedge \cF - \frac{1}{4} \tilde{\lambda} \wedge \cG \eol
	&= \imath_\xi \cH
		- \frac{1}{4} \cW \wedge \imath_\xi \cF - \frac{1}{4} \tilde{\cW} \wedge \imath_\xi \cG~.
\end{align}
This can be rewritten using $\delta^{\rm cov}_\xi \cW = \imath_\xi \cF$
and $\delta^{\rm cov}_\xi \tilde{\cW} = \imath_\xi \cG$ as
\begin{align}
\delta^{\rm cov}_\xi \cB
	&= - \frac{1}{4} \cW \wedge \delta^{\rm cov}_\xi \cW -\frac{1}{4} \tilde{\cW} \wedge \delta^{\rm cov}_\xi \tilde{\cW} + \imath_\xi \cH~.
\end{align}
Projecting to $\q=\rd\theta=0$ exactly reproduces \eqref{varC}.

\section{Conclusion and future directions}
In this paper, we constructed the dilaton Weyl multiplet in four dimensions. This differs from the standard Weyl multiplet in its auxiliary field content. It contains two additional U(1) symmetries and one vector-gauge transformation which are absent in the standard Weyl multiplet. The existence of two different versions of the Weyl multiplet was already known in five and six dimensions \cite{Bergshoeff:2001hc,Fujita:2001kv,Bergshoeff:1985mz}. In this paper we have filled the gap by explicitly constructing the dilaton Weyl multiplet in four dimensions. We have also shown its equivalence to the multiplets constructed in \cite{Muller:1986ku,Siegel:1995px} using superspace. 

In principle, we can also obtain this multiplet from the five dimensional dilaton Weyl multiplet \cite{Bergshoeff:2001hc,Fujita:2001kv} by dimensionally reducing it on a circle using the procedure outlined in \cite{Banerjee:2011ts}. This is currently a work in progress. 

Using the standard Weyl multiplet in four dimensions, a higher derivative invariant based on the logarithm of a conformal primary chiral superfield was constructed in \cite{Butter:2013lta}. A particular linear combination of this with the well-known Weyl squared invariant \cite{Bergshoeff:1980is} gives the supersymmetrization of the Gauss-Bonnet term. A separate supersymmetric $R^2$ invariant was constructed in \cite{Kuzenko:2015jxa}. This gives all three possible curvature squared invariants in 4d $N=2$ supergravity using the standard Weyl multiplet.
It would be interesting to show, in complete analogy to the five dimensional case 
\cite{Hanaki:2006pj, Ozkan:2013nwa, Butter:2014xxa}, that all curvature squared invariants can also be obtained in the 4d $N=2$ dilaton Weyl formulation.
These higher derivative invariants could in principle be obtained either by dimensional reduction from the 5d dilaton Weyl invariants \cite{Ozkan:2013nwa} or directly in the four dimensional context. The former procedure would have the important advantage of establishing useful explicit relations between the curvature squared invariants in five and four dimensions, similar to what was shown in \cite{Banerjee:2011ts, Butter:2014iwa}.
Furthermore, it would be interesting to see the relevance of the
dilaton Weyl multiplet and the new invariants in the context of
black hole entropy, including for non-BPS and small black holes,
and possibly use it to understand the puzzles involving the entropy
of five dimensional black holes given in \cite{Cvitan:2007hu}. We leave this for future work.

\acknowledgments{
The work of D.B. is partially supported by NSF under grants PHY-1521099 and PHY-1620742 and
the Mitchell Institute for Fundamental Physics and Astronomy at Texas A\&M University.
}

\bibliography{references}
\bibliographystyle{jhep}

\end{document}